\documentclass[11pt,twoside]{article}
\usepackage{asp2004}
\usepackage{psfig}
\usepackage{epsf}
\def\edcomment#1{\iffalse\marginpar{\raggedright\sl#1\/}\else\relax\fi}
\reversemarginpar

\begin{document}
\title{The HI Parkes Deep Zone of Avoidance Survey }
\author{Patricia A. Henning}
\affil{Institute for Astrophysics, University of New Mexico, 800 Yale Blvd.,
NE, Albuquerque, NM, 87131, USA}
\author{Ren\'ee C. Kraan-Korteweg}
\affil{Depto.~de Astronom\'ia, Universidad de Guanajuato, Apdo.~Postal
144, Guanajuato, GTO 36000, M\'exico}
\author{Lister Staveley-Smith}
\affil{Australia Telescope National Facility, CSIRO, P.~O.~Box 76,
Epping, NSW 1710, Australia}

\begin{abstract} The 64-m Parkes telescope, equipped with the 21-cm 
multibeam receiver,
has completed a sensitive survey (typically 6 mJy beam$^{-1}$ rms) 
for HI galaxies in the Zone of Avoidance (ZOA) accessible to the telescope, 
$196^\circ \le \ell \le 52^\circ$, and $|b|$~$\leq$~5$^\circ$.
While galaxy candidate inspection is not yet quite complete, and final
number not yet determined, the survey has yielded about 1000 galaxies.
The data, in the form of three-dimensional datacubes, have been inspected
by eye, and candidate lists assembled, and about half have now been
checked for reality, and accepted into the final catalog.  
The distributions on the sky and in redshift space are presented,
showing galaxies belonging to previously-known structures, and newly-discovered
features. 
Of the 469 confirmed HI galaxies, 191 have a NIR
source within $6\arcmin$ in the 2MASS Extended Source Catalog, but the
incidence of NIR counterparts is a strong function of longitude:
in the low obscuration, low stellar surface density
Puppis region, 131 of the 186 HI galaxies have
2MASS counterparts (70\%), while in the Galactic bulge region, only 6 of the
155 HI detections have a 2MASS extended source coincident (4\%).  
This is attributable
to the HI survey's ability to detect galaxies even in regions of high
foreground stellar surface density.

\end{abstract}
\thispagestyle{plain}

\section{Why a 21-cm Survey?}

Dedicated searches for galaxies and clusters have been
successful in recent years in narrowing the 
ZOA, and obtaining redshifts, where possible, to map 
three dimensional large-scale structures at low Galactic latitude
(see Kraan-Korteweg \& Lahav 2000 for an overview of the various
multiwavelength campaigns in the ZOA, and the many contributions in 
this volume.)
Results from the 2 Micron All-Sky Survey (2MASS) have produced an 
impressive narrowing of the ZOA, particularly away from the Galactic 
bulge region,
and allowed appreciation of low-latitude large-scale structure traced
by a homogeneous sample (see contribution from Huchra in
this volume.)  
However, such methods fail in regions of
heaviest obscuration, and highest foreground stellar surface density.
Fortunately, 21-cm searches for HI-bearing galaxies have been proven to
succeed in 
obscured and confused areas (eg. Kerr \& Henning 1987; Kraan-Korteweg 
et al.~1994; Henning et al.~1998; Henning et al.~2000.)

\section{The Survey}

\subsection{Parameters and Strategy}

The multibeam receiver system on the Parkes 64-m telescope, with
its large footprint on the sky, allows a sensitive,
wide area survey.  The ZOA accessible from Parkes, covering 
$196^\circ \le \ell \le 52^\circ$, $|b|$~$\leq$~5$^\circ$, was observed
to quite uniform sensitivity, due to the strategy of observing
overlapping strips of constant Galactic latitude
(Staveley-Smith et al.~1998; Henning et al.~2000; Donley et al.~2004).  
The data were
bandpass-corrected, Doppler corrected, calibrated and gridded with 
resulting pixel and beam sizes of $4\arcmin \times 4\arcmin$, and
$15\farcm5$, respectively.  The observations were done over 27 fields of
($\Delta \ell, \Delta b = 8^\circ, 10^\circ$).  The effective integration
time per beam was 2100 s [compare with 450 s per beam integration time of
the HI Parkes All Sky Survey (HIPASS; Meyer et al.~2004)]. 
The correlator bandwidth
of 64 MHz, set to cover the velocity range $-1200$ to 12700 km s$^{-1}$,
provides coverage in the third dimension, thus we speak of the 
{\it datacubes}, three-dimensional position-position-velocity representations
of the survey data.

Because of the strong HI signal of the Galaxy, which causes
spectral ringing, the data were Hanning smoothed, resulting in a
velocity resolution of 27.0 km~s$^{-1}$, a significant increase over
the channel spacing of 13.2 km~s$^{-1}$.   
Strong continuum emission was subtracted,
although some residual continuum baseline ripple
remains where there was particularly strong continuum emission.  

\subsection{Survey Sensitivity}

Over most of the volume surveyed, away from small regions of strong continuum,
the noise was about 6 mJy beam$^{-1}$ rms, which compares
favorably to the HIPASS noise of 13 mJy beam$^{-1}$.
This noise of 6 mJy beam$^{-1}$ is equivalent to a
5$\sigma$ HI mass detection limit of $1.4 \times 10^6$ d$^2\!\!_{\rm Mpc}$
M$_{\odot}$ (for a galaxy with a linewidth of
200 km~s$^{-1}$).  For instance, the survey was sensitive to galaxies with 
$5 \times
10^9$ M$_{\odot}$ at 60 Mpc, and $1 \times 10^{10}$ M$_{\odot}$ at 100 Mpc. 
Thus, the survey was sensitive to normal spirals well beyond the Great Attractor
region, and could detect some galaxies beyond 10,000 km~s$^{-1}$.
It was also sensitive to very low mass local dwarfs, as well.

\subsection{Searching for Galaxies}

The 27 datacubes were searched by members of the HI Parkes ZOA
Team.  The cubes were searched by eye, because experimentation with 
automatic galaxy detection algorithms
indicated that in the complicated region of the ZOA, with regions of
increased noise due to continuum sources and Galactic HI, the human
eye-brain system is enormously more effective at finding galaxies.
Each cube was searched by two, or sometimes three, independent
searchers, using the visualization tool \verb"karma KVIEW" (Gooch 1995),
producing independent lists of galaxy candidates.  
Figure 1 shows two slices through one of the datacubes, as a searcher
would display and search the data.
While the entire velocity
range of the data was searched, confusion due to Galactic HI generally
prevented the recognition of galaxies within about $|v| \leq 250$ km~s$^{-1}$.
In addition, the higher noise at low latitudes near the Galactic Center
creates a residual ZOA, but even quite close to the Galactic Center direction
we were able to detect galaxies (Fig.~2).

\begin{figure}[!ht]

\plottwo{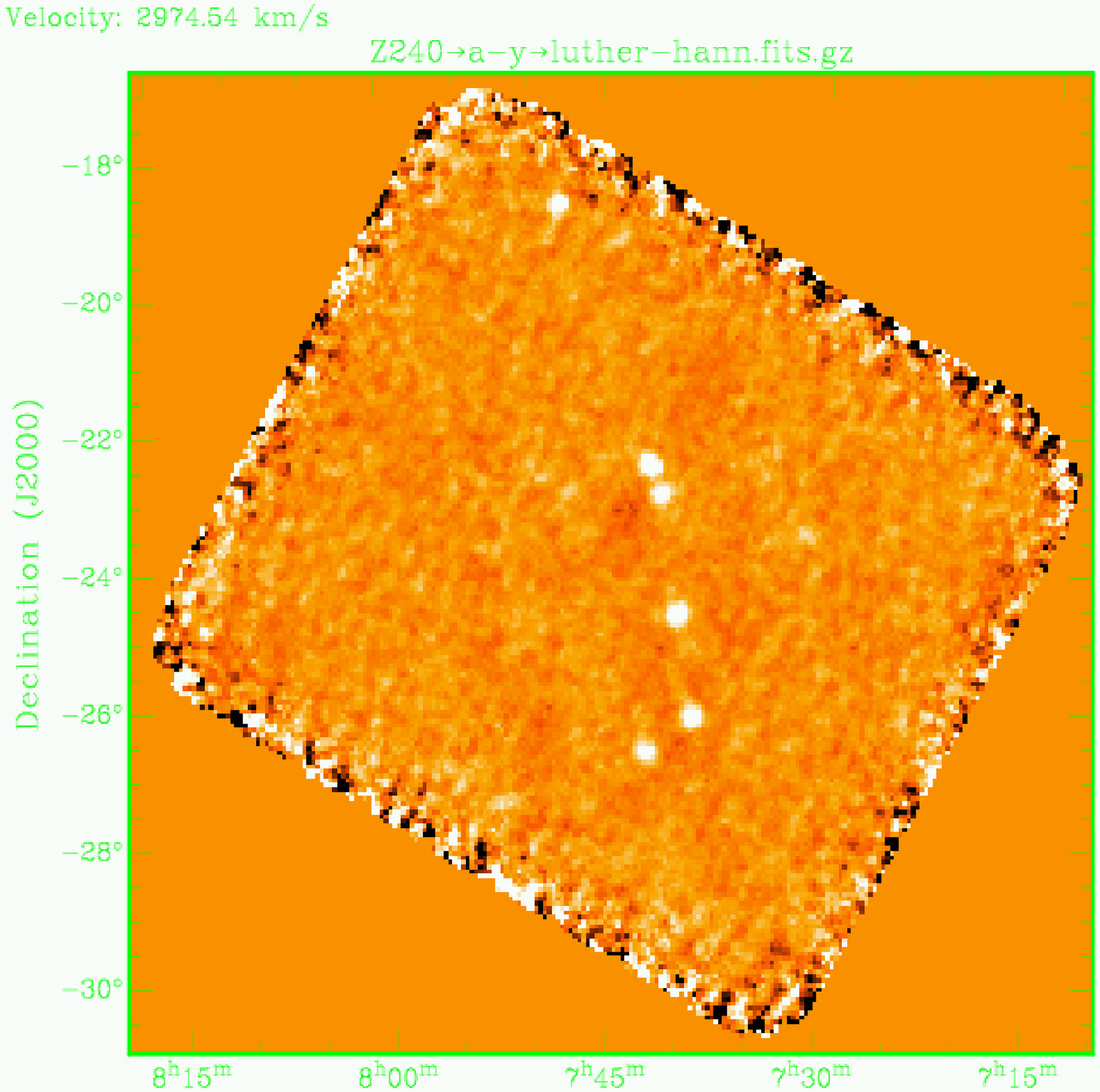}{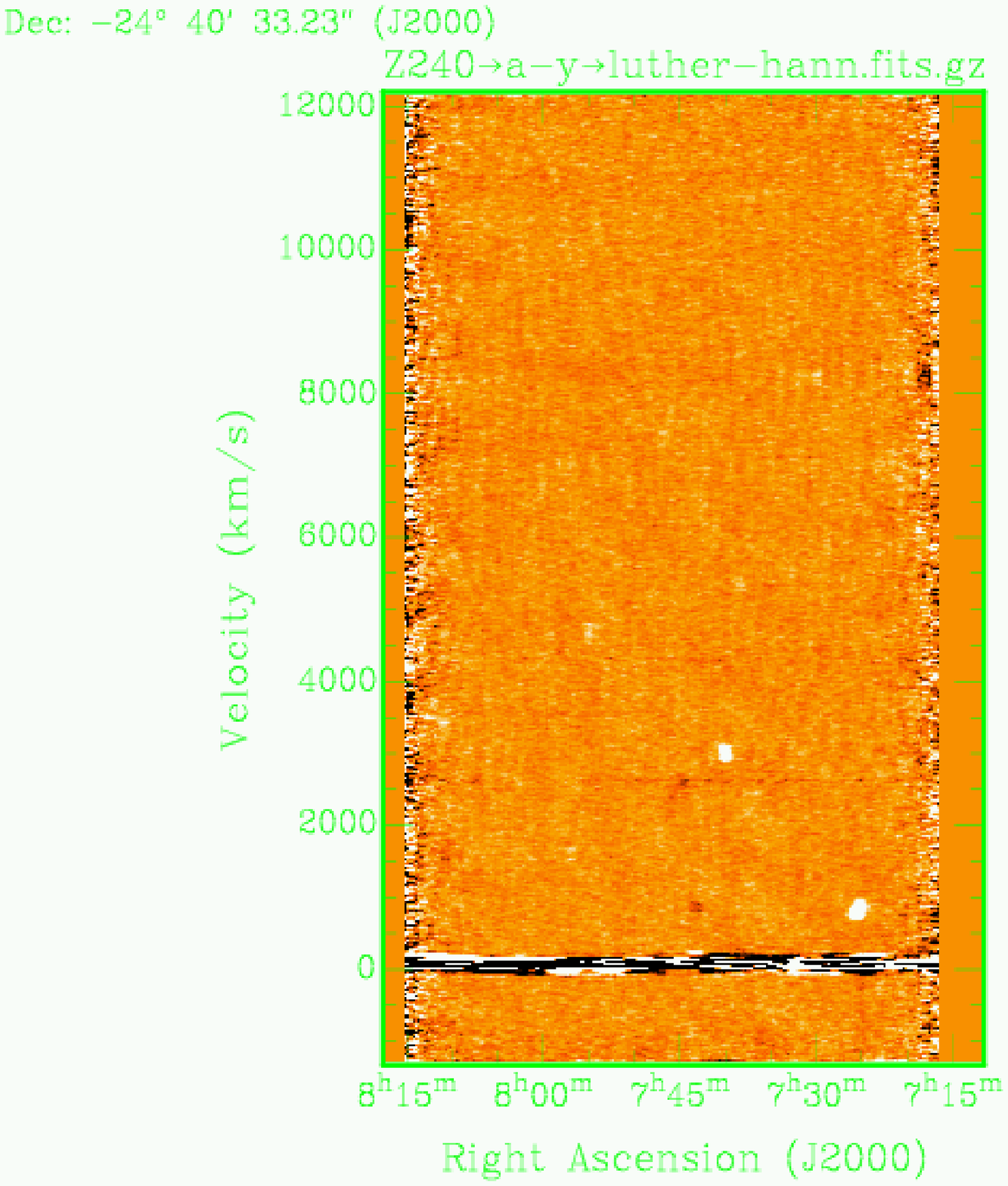}

\caption{Left panel shows a right ascension -- declination plane
of a datacube, in a single velocity channel.  Several galaxies are visible
as white blobs in this rich region.
The right panel shows a right ascension -- velocity plane, at the declination
of one of the galaxies appearing in the left panel.   In addition to the
extragalactic sources,
Galactic HI appears as the strong horizontal feature at zero velocity.}

\end{figure}

\begin{figure}[!ht]

\plottwo{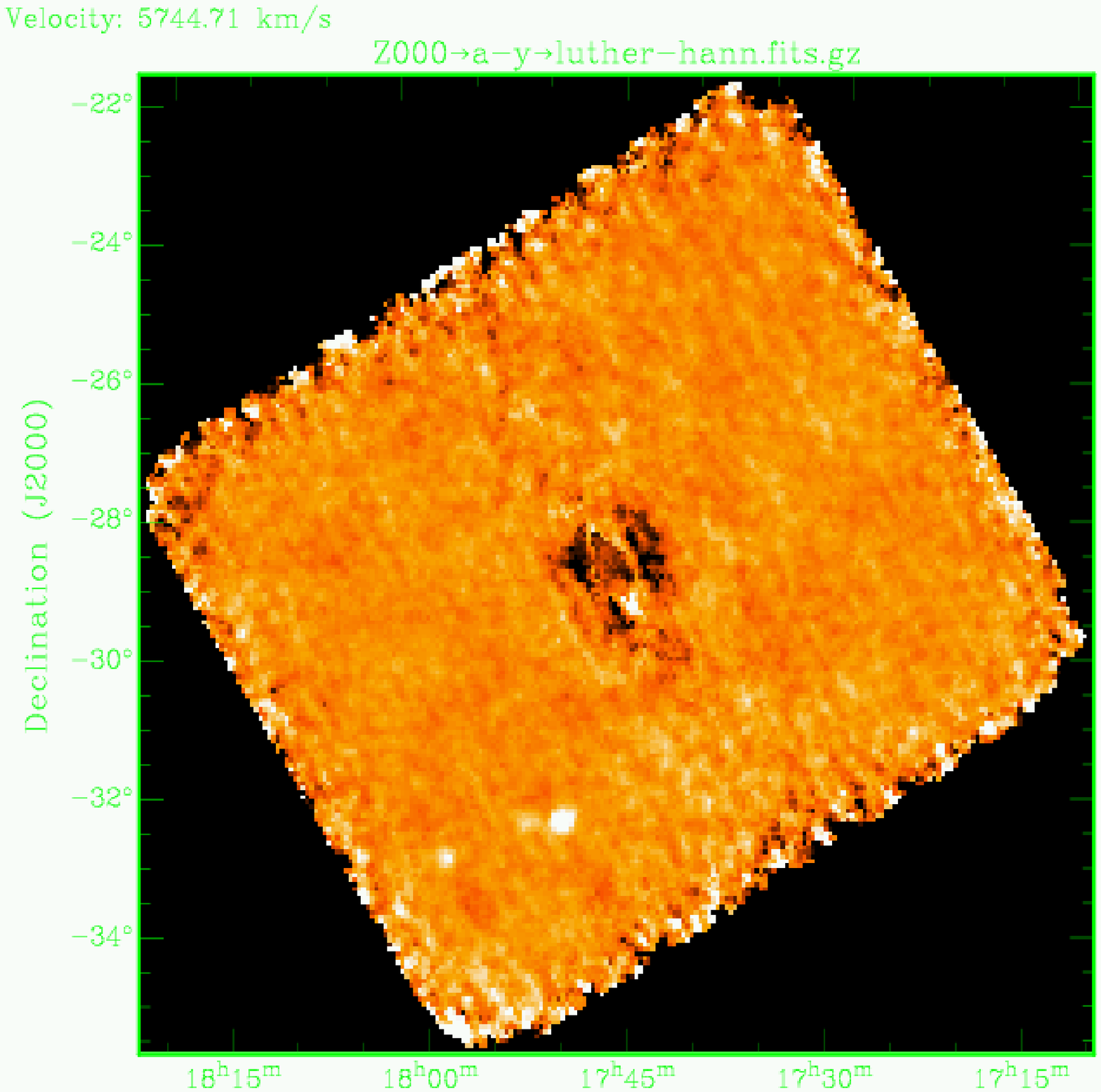}{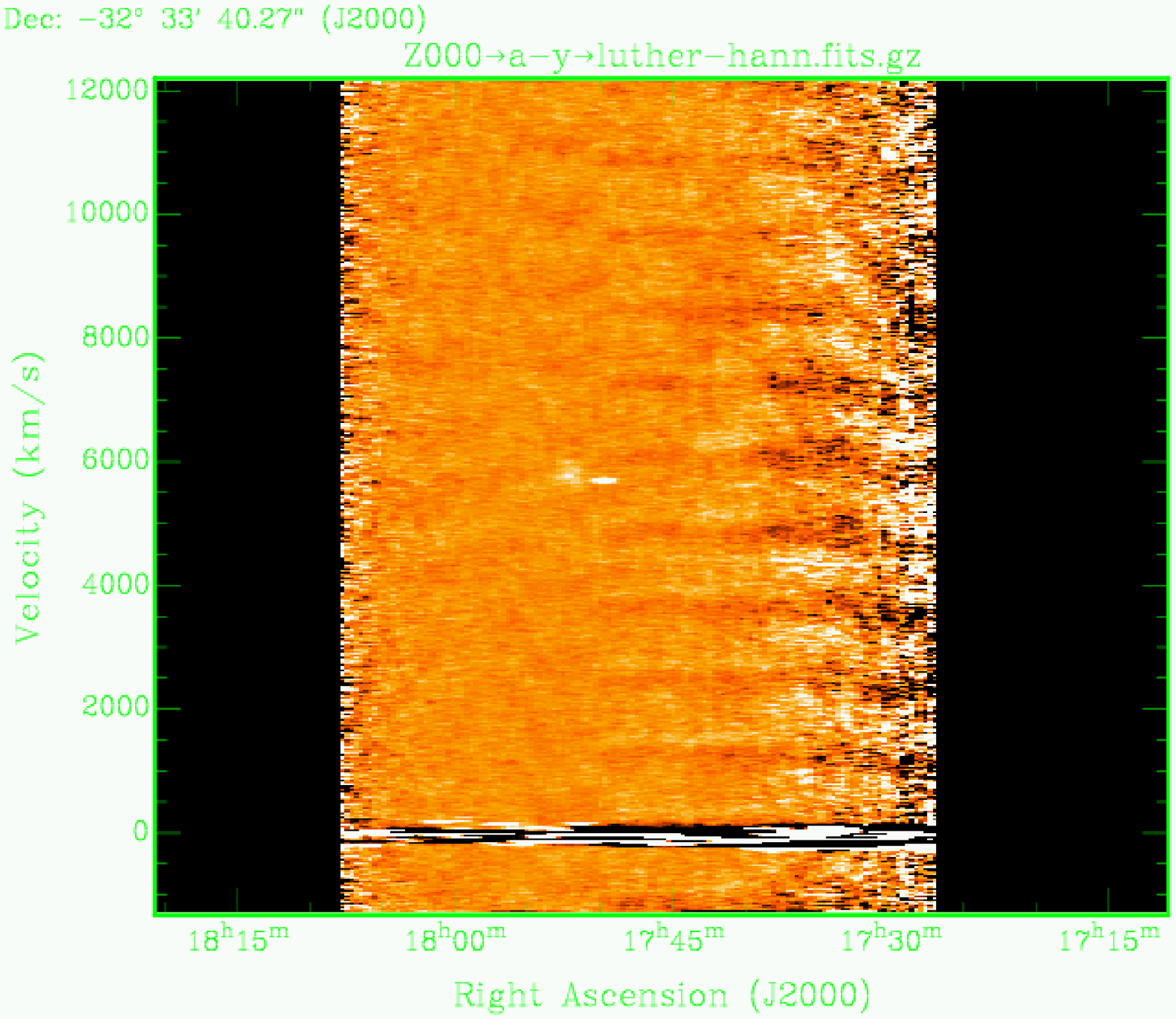}

\caption{Left panel shows a right ascension -- declination slice which
contains the Galactic Center, visible as noisy, disturbed data at the
center of the map.  Note the 3 extragalactic sources visible in this plane.
The right panel shows the right ascension -- velocity slice through two
of the sources at the same declination on the left-hand map.  Despite the
increased noise, these galaxies are clearly detected within a few degrees
of the Galactic Center position itself.}

\end{figure}

The datacubes have all been searched, and a list of over 
1000 galaxy candidates has been produced.
To generate a uniformly-selected catalog, all galaxy candidates will be
inspected by one person (PAH), who decides if a candidate 
is to be included in
the final catalog.  As of writing (June 2004), 16 of the 27 cubes
have been inspected by the adjudicator, with 469 candidates accepted.
The adjudication is complete over the longitude 
ranges of $332^\circ$ to $52^\circ$, 
$236^\circ$ to $260^\circ$, and other isolated
longitudes.  In the regions $\ell = 36^\circ$ to $52^\circ$ and $196^\circ$ 
to $212^\circ$ (the ``northern extension''), 
the data are presented by Donley et al.~(2004).

\section{Distribution of the Detected Galaxies and Candidates}

In Figure 3, the upper panel shows the distribution of galaxies in the 
literature (selected
from the LEDA database), and the outline of our survey region in the ZOA.
The traditional ZOA is clearly visible as a paucity of galaxies at low
Galactic latitude.  In the lower panel, with our confirmed 
galaxies and candidates
added in, the ZOA fills in remarkably well, and several large-scale
structures are seen to cross the Galactic Plane.  The Great Attractor
region is seen as an overdensity at $\ell \sim 300^\circ \rightarrow 
340^\circ$, the Hydra-Antlia filament at $\ell \sim 280^\circ$
and the Puppis filament crosses at $\ell \sim  240^\circ$.  The Local
and Sagittarius voids are visible as an underdensity of galaxies at
$\ell \sim 350^\circ \rightarrow 52^\circ$.    

\begin{figure}[!ht]

\plotone{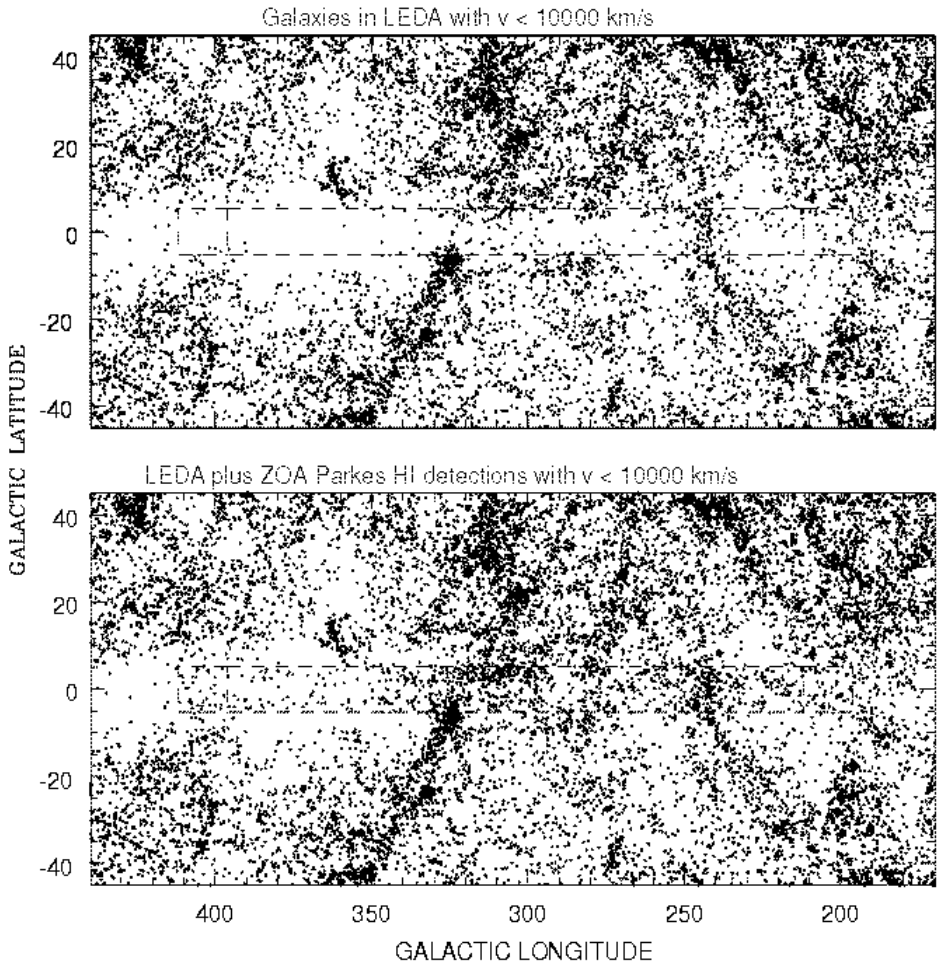}

\caption{Top panel shows galaxies from the LEDA database with velocities
measured within 10000 km~s$^{-1}$.  The dashed black lines show the
borders of the ZOA search area in the southern hemisphere, and extension 
to the north.  The lower
panel adds in the galaxies and candidates uncovered by the HI Parkes
Deep ZOA Survey.}

\end{figure}

In velocity space, as shown in Figure 4, quite a number of structures become
apparent.  The broad overdensity of the Great Attractor is evident at
velocity $\sim 5000$ km~s$^{-1}$.  Only a portion of this feature was
known from higher-latitude catalogs, and is labelled as part of the Norma
supercluster in Figure 5.  Galaxies associated with the PKS 1343-601
cluster are also labelled in Figure 5, as is the Puppis cluster,
and background void.  The filament hinted at, at $\ell \sim 220^\circ$ 
in Figure 3 becomes clear in the wedge diagram, which we label ``Hydra wall
and Monoceros extension'' in Figure 5.  

\begin{figure}[!ht]

\plotone{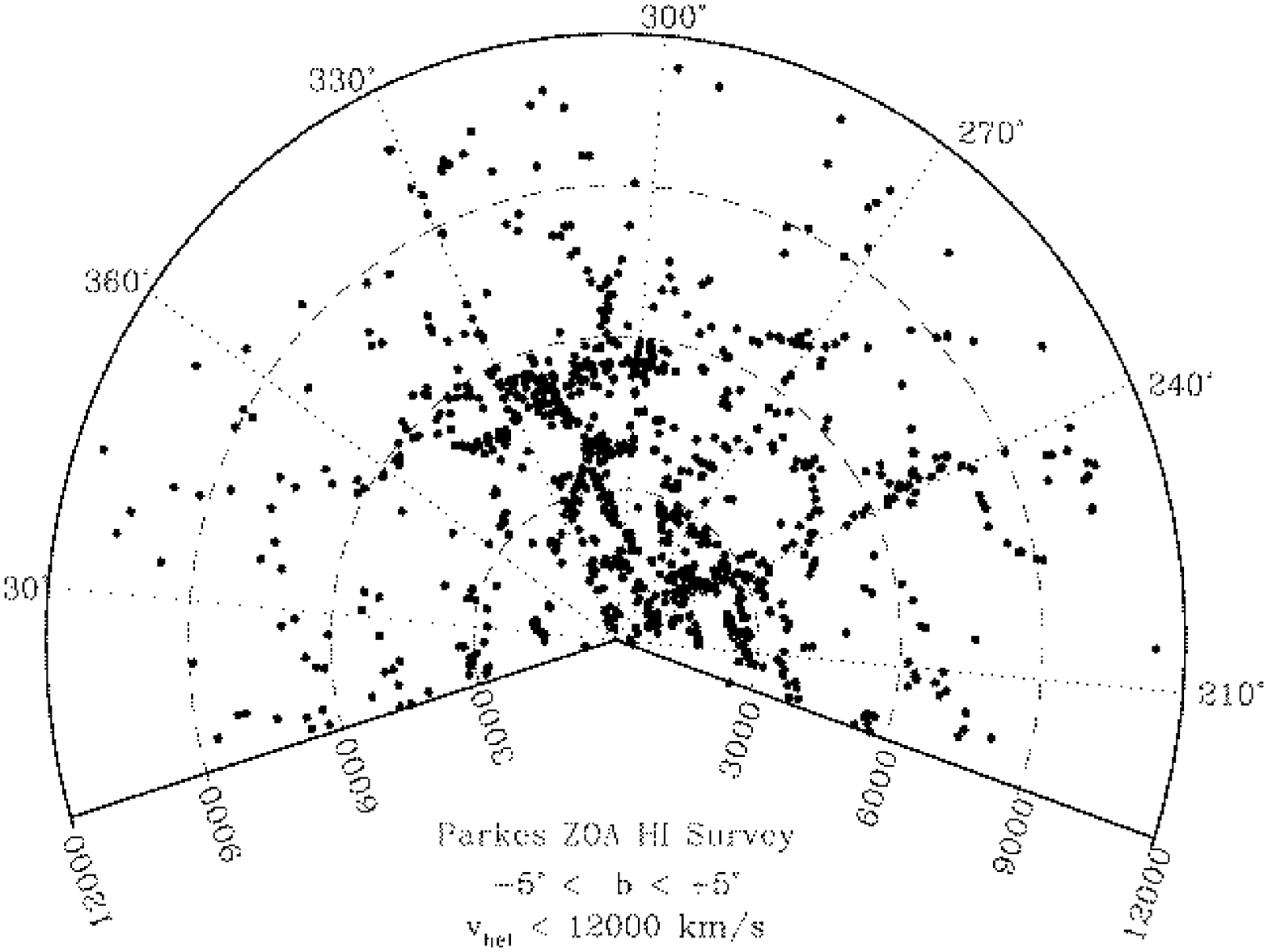}

\caption{Distribution of HI sources in Galactic longitude and recessional 
velocity.
Note that sources are detected beyond 
10000 km~s$^{-1}$.  This figure is very similar to Kraan-Korteweg et al.'s 
Figure 3 to be published in the IAU Symp.~Vol.~216, but with some spurious 
sources in the Local Void
area removed.  This figure, too, is provisional, until all candidates have been
examined, but we expect little qualitative change.}

\end{figure}

\section{2MASS Counterparts}

Of the 469 confirmed HI galaxies (most others will certainly be real,
but the adjudication is not yet complete), there are 191 for which 
there is a NIR source listed
within 6\arcmin in the 2MASS Extended Source Catalog.  We have not yet
examined these NIR sources to check if they are all galaxian, or if some
are confused
Galactic objects, quite possible at low latitudes.  In the longitude range
$236^\circ$ to $260^\circ$, there are 186 HI detections, 131 of which have
an extended 2MASS counterpart (70\%).  As we continue to study the HI-selected
galaxies, we can investigate the populations of ZOA galaxies uncovered at
21 cm {\it vs.~}at NIR wavelengths.  One would expect the HI galaxies to
be generally of later type, but this will be quantified better in the future.

In contrast, in the Galactic bulge 
region, $\ell = 332^\circ$ to $36^\circ$, only 6 of the 155 HI detections have
a 2MASS extended source within 6\arcmin of the HI (4\%).  The poor performance
of the NIR survey in recovering the HI-selected galaxies is due to the
extreme stellar confusion in the Galactic bulge.  The completeness of
the 2MASS survey is a function of stellar survey density, and drops even
as far as $\ell \pm 90^\circ$ from the Galactic Center near the Galactic
plane.

\begin{figure}[!ht]

\plotone{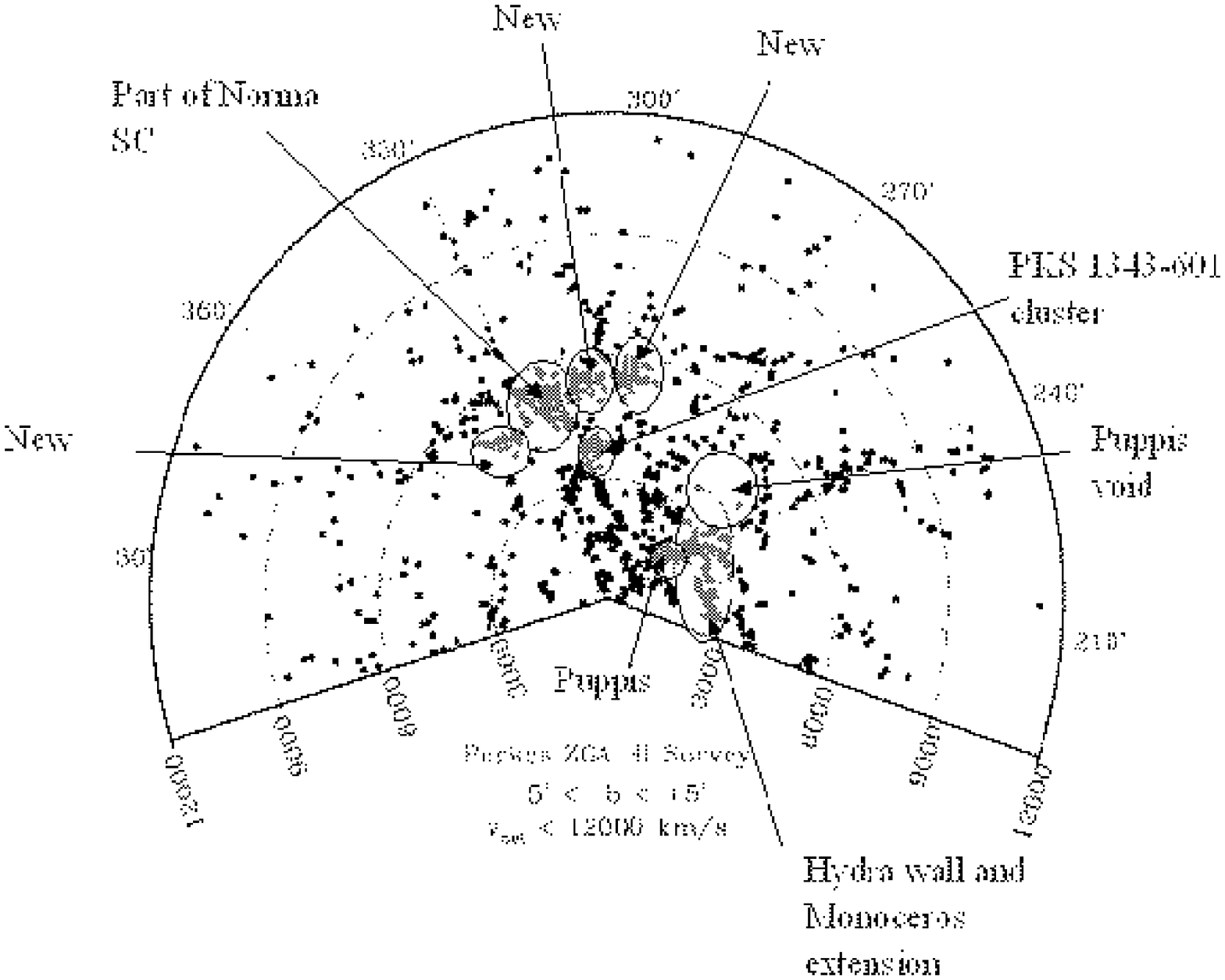}

\caption{As in Figure 4, but with some new, and previously-indicated
large-scale structures labelled.}

\end{figure}

\section{Ongoing Work}

To produce the final catalog of the HI Parkes Deep ZOA Survey, we are working
to finish the adjudication of candidates.  
We then will measure HI parameters,
and quantify the selection function {\it a posteriori}, since this 
sample has been selected by human eyes.  When the full sample is constructed,
its HI properties, large-scale structures traced, and NIR counterparts will
be investigated.
We are also extending the search region with new observations 
to higher latitudes in the Galactic
bulge area, since there the ZOA extends further than $\pm5^\circ$ from 
the plane (see Fig.~3.)
Finally, we are exploring the possibility of using the new 21-cm multibeam
receiver at the Arecibo Observatory to extend 
our map of hidden HI galaxies and large-scale structures further to the
north.

\acknowledgments{We very warmly thank the other ZOA cube searchers 
(J.~Donley, A.~J.~Green, S.~Juraszek, B.~S.~Koribalski, E.~M.~Sadler,
I.~Stewart, and A.~Schr\"oder) and other participants in the survey.
This research used the Lyon-Meudon Extragalactic Database (LEDA),
supplied by the LEDA team at the Centre de Recherche Astronomique
de Lyon, Obs.~de Lyon.  We also made use of the NASA/IPAC Extragalactic
Database, which is operated by the Jet Propulsion Laboratory, California
Institute of Technology, under contract with the National Aeronautics and
Space Administration.  PAH is grateful for the support of 
the American Astronomical Society and the National Science Foundation 
in the form of an International Travel Grant, which enabled her to
attend this conference.  
RCKK thanks CONACyT for their support (research grants 27602E and
40094F).
Finally, we thank our Cape Town hosts, 
Prof.~Anthony Fairall and Dr.~Patrick Woudt for creating a wonderful
atmosphere at the conference for science, and for showing us beautiful
Cape Town and the Peninsula.}


\begin{references}

\reference Donley, J.L., Staveley-Smith, L., Kraan-Korteweg, R.C.
et al.~\aj, submitted

\reference Gooch, R.E. 1995, in ASP Conf.~Ser.~101, Astronomical Data Analysis
Software and Systems V, ed.~G.H.~Jacoby \& J.~Barnes (San Francisco:  ASP),
80

\reference Henning, P.A., Kraan-Korteweg, R.C., Rivers, A.J. et al.~1998,
\aj,115, 584

\reference Henning, P.A., Staveley-Smith, L., Ekers, R.D. et al.~2000,
\aj, 119, 2686

\reference Kerr, F.J., \& Henning, P.A.~1987, \apj, 320, L99

\reference Kraan-Korteweg, R.C., Loan, A.J., Burton, W.B. et al.~1994, \nat,
372, 77

\reference Kraan-Korteweg, R.C., \& Lahav, O.~2000, A\&ARv, 10, 211

\reference Kraan-Korteweg, R.C., Staveley-Smith, L., Donley, J., 
\& Henning, P.A. in IAU Symp.~216, Maps of the Cosmos, eds.~Matthew Colless \&
Lister Staveley-Smith, (San Francisco: ASP), in press (astro-ph/0311129)

\reference Meyer, M.J., Zwaan, M.A., Webster, R.L. et al.~\mnras, 350, 1195

\reference Staveley-Smith, L., Juraszek, S., Koribalski, B.S.~ et al.~1998,
\aj, 116, 2717

\end{references}
\end{document}